# A Systematic Method of GPU Verification on Subsystem Modelling

Jiangning Liu (jiangning.liu@arm.com), Kevin Qin (kevin.qin@arm.com),
Patrick Zhang (patrick.zhang@arm.com), Hale Wang (hale.wang@arm.com),
Robert Catherall (robert.catherall@arm.com)

**Abstract - ARM Fast Model is a software modelling system for simulating multiple systems IPs integrated in a subsystem. Fast Model allows customer to run whole software stack as early as system IPs' specification is just ready. In this paper, we talk about a systematic method to enable GPU modelling and make whole system software stacks executable for both CPU and GPU in the integrated modelling environment, and eventually the major GPU IP can be verified. With our solution, we can directly run commercial benchmarks like Antutu and GFX Bench in Android target system with underlying ARM Fast Model mobile subsystem support. This was unbelievable in the past due to 3D rendering simulation for GPU verification is very slow. We also introduce some special techniques to improve performance. Eventually the whole software stack can directly work on real hardware without any changes as soon as the first tape-out is ready.**

**Keywords --- GPU verification, OpenGL ES, Android, Modelling, Simulation**

## I. BACKGROUND INTRODUCTION

ARM Fast Model [1] is an environment for easy creation of virtual platform models that execute with high simulation speeds. ARM hardware IPs interfaces can be defined by LISA+ code and easily visualized and combined to generate a complete subsystem. Modern SoC vendors tend to provide whole software stack support along with the first chip tape-out ready. In ARM Fast Model generated platforms, we can run not only bare-metal benchmarks but also mainstream mobile operating systems like Android and other application software running on CPU. Apart from this, it becomes more interesting in verifying GPU IP capability because of the developed mobile ecosystem in recent years.

GPU architecture is evolving quickly, and it tends to use very big number of unified scalar cores to solve highly data parallel stream processing problem. Compared with GPU, typically CPU has fewer cores, but a single CPU core is more complicated than a single GPU core, e.g. deeper pipeline and bigger caches. For history reasons, the GPU is always treated as a peripheral device in CPU centric computer architecture. For whole system simulation, usually we use CPU to simulate all devices including 3D rendering using GPU, but the big architecture level difference between CPU and GPU makes fast GPU simulation impossible. The increasing display resolution in current mobile system is demanding even higher computation resources for simulation environment.

Besides, in industry the trend is to fuse GPU with CPU to solve the nagging issue about memory access capability and bandwidth limitation between CPU and GPU. Anyway, it would be a great challenge to use limited CPU resources to simulate and verify GPU IP. The simulation and verification speed of whole system could be very slow like days or months to run a simple workload.

To make CPU architecture study work effectively, some researching works, e.g. Gem5 [4] NoMali Model [9], implicitly ignore real GPU workloads to make the application using 3D rendering on GPU can still run. This way GPU memory system interaction is not simulated completely, so it cannot visualize applications with rich graphics features, and bring trouble to real Android software stack enabling work in reality. To visualize application with simple graphics like 2D GUI on screen, Google Android Emulator uses QEMU [6] to do software rendering, which is essentially to use CPU resource to simulate simple graphics work. With this solution, complicated 3D rendering as defined by GLES 2.0 [3] and above cannot really be supported, so end users can't run 3D benchmarks like Antutu, GFX and unity.

Traditionally graphics researches use tracing technology to capture all run-time GL/GLES APIs calls and parameters, and get them replayed offline by using either Cycle Models [8] or C model. People uses Cycle Model to study GPU architecture, but it can only run several frames due to extremely slow speed. C model can be a little bit faster, but GPU study is the only focus, and the computation in CPU side computation is completely ignored at trace replay time. This can't help the interleaved logic on CPU and GPU in practical benchmarks. Besides, sometimes C model is not fast enough to run commercial benchmarks.

However, GPU simulation by using GPU hardware resources in host could make the combined GPU/CPU simulation in target possible. In industry, the software programming model for 3D rendering using GPU has evolved to be using consistent APIs, e.g. Open GL on Linux, and DirectX on Windows, although the hardware architecture for different generations of GPUs may be significantly different. It would be much easier to accelerate 3D rendering in target GPU by passing them to run on host GPU directly.

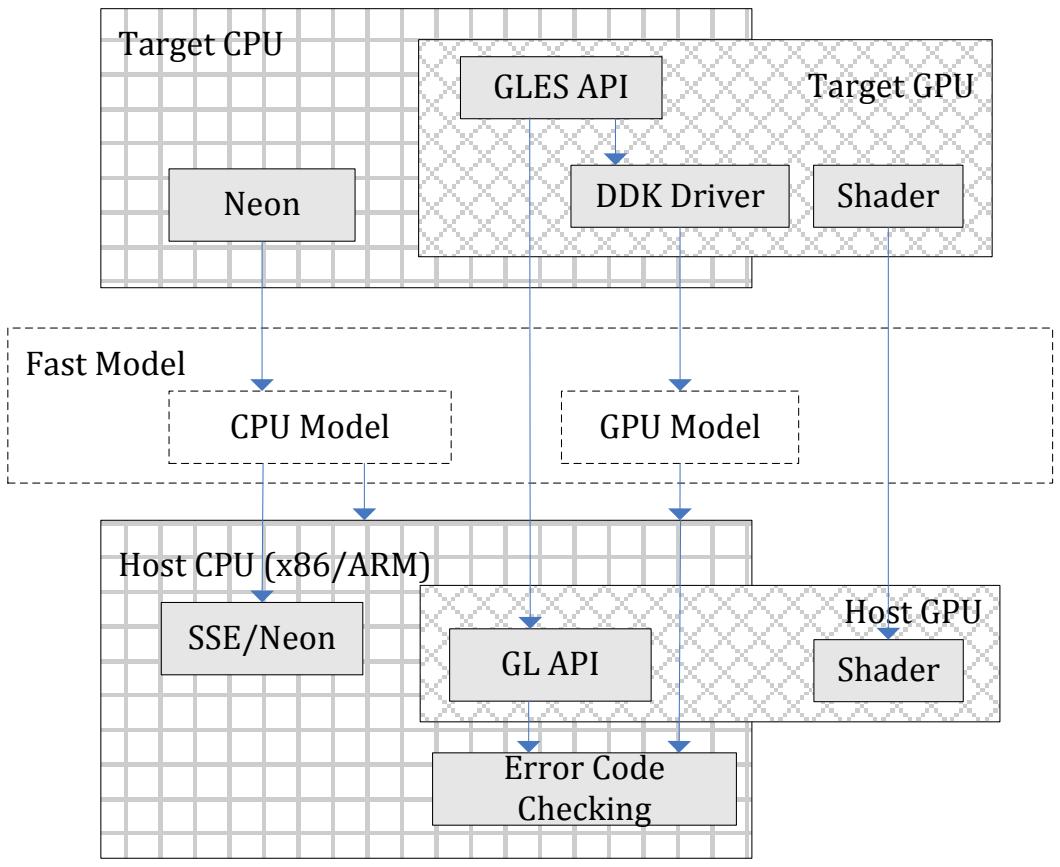

Figure 1. Fast Model 3D rendering acceleration

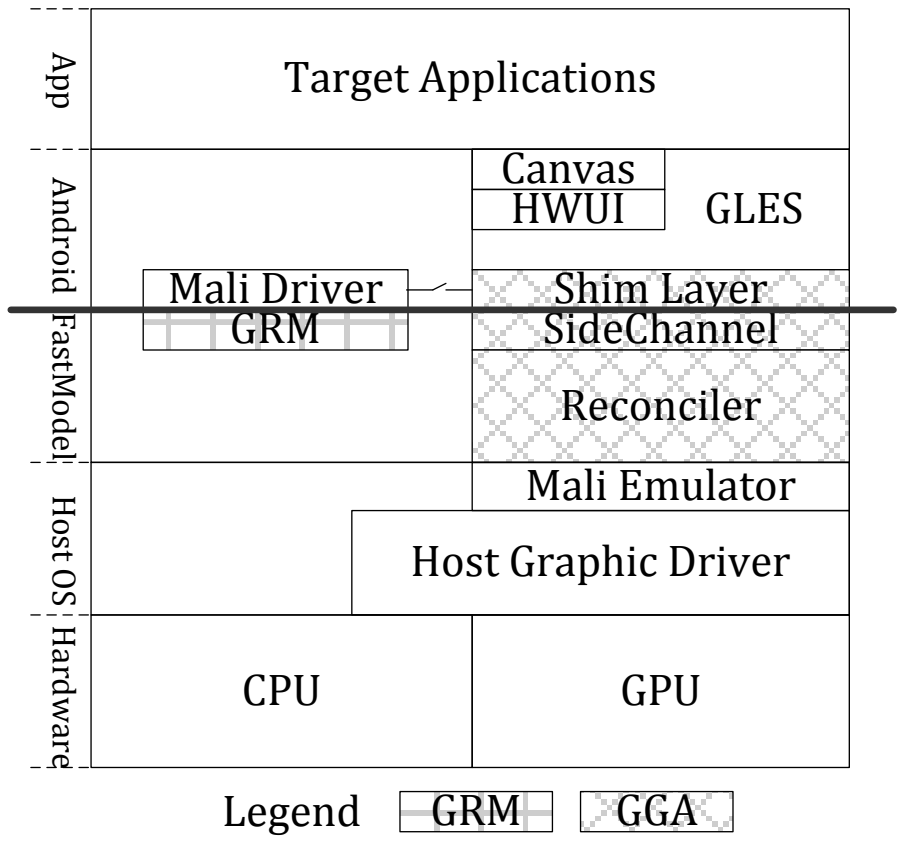

Figure 2. Android Graphics Software Stack

Figure 1 illustrates how Fast Model support 3D rendering acceleration in a complete run-time working environment. Target and host are completely isolated by Fast Model. The CPU computation in target can be directly translated into instructions in host, which is usually a cross architecture translation. The driver in target to support GPU also runs on CPU, but the shader code in target runs on GPU. The EGL/GLES APIs are directly dispatched to host and GPU in host does real 3D rendering work accordingly. In the meantime, we have chances to compare the error code returned by target and host at verification stage in host, which is very important to help GPU driver correctness verification before running on real GPU hardware.

In this paper, we propose a systematic framework to make full graphics software stack run on a subsystem of ARM Fast Model. With this framework, user level applications, native graphics drivers in both user space and kernel space, Android HWUI, Surface Flinger, Gralloc HAL [12] and others related to graphics can all run in a monolithic Fast Model virtual environment. With this approach we can not only easily detect all of the system integration issues, but also expose memory traffics between CPU and GPU. This give us opportunity to capture the graphics driver bugs as well as performance bottleneck around memory traffic introduced by GLES APIs. We also propose an enhanced solution to support remote 3D rendering acceleration and this solution eventually helps the real GPU verification work running on massive cluster machines in a real production environment.

Compared with GPU-HIL simulation [7] technique proposed by Youngsub et al., our proposed work has two major advantages (a) Support graphics register model, so the graphics driver in target can be directly executed and verified. (b) Zero Android image change at tape-out time, so most of the integration issues can be captured at the earliest software development time.

The rest of the paper is organized as following. Section II gives an overview of 3D rendering acceleration using host GPU for Android OS. Section III describes the potential capability of GPU driver verification supported by GPU model. Section IV proposes the remote 3D rendering solution for cluster. Section V is the experimental result of graphics software stack simulation performance. Section VI talks about the conclusion and the work in future.

## II. RENDERING ACCELERATION USING HOST GPU

In order to accelerate the GPU IP modelling in ARM Fast Model, we explored a new method by dispatching the computation from target GPU to host GPU. Graphics software stack is very different from the software ecosystem in CPU side. Traditionally graphics end user uses high level APIs, e.g. DirectX or Open GL/GLES, to help 3D rendering development. The real workload is dispatched to GPU tiny cores as so-called shader or kernel. (Figure 1) It's very challenge to translate the shader binary code from target GPU ISA to a host GPU one, and it would be very time consuming as well. Therefore, we dispatch shader source or IL instead, and ask GPU in host side to do real rendering work, and then push framebuffer data back to targeting Android OS running in Fast Model. This way, we can fully enable 3D rendering in Fast Model platform, and end user can view the 3D rendering result directly in virtualized Display devices. This is a very attractive and realistic feature for SoC software stack developer to make sure hardware and software development can really be done in parallel.

### 2.1 OpenGL ES in Android

Figure 2 shows the graphics software stack in Android OS. ARM Fast Model isolated Android OS target from host OS. In Android target, OpenGL ES APIs could be invoked by a couple of different sources including end user application, Android home launcher, HWUI invoked by Canvas, and Surface Flinger etc. Apart from a very thin layer of EGL/GLES wrapper, OpenGL ES APIs can be intercepted by libraries in Android. In our approach, we use a library called Shim Layer (libShim.so) in vendor folder to intercept all APIs and dispatch them to both graphics drivers in target and the Reconciler in host. On one hand, the native graphics driver in target runs on a NoMali model, which is shared with the implementation in Gem5. The difference from Gem5 solution is for each API we need to wait in target until the API is really dispatched to host. On the other hand, the API is passed to host graphics driver to complete real 3D rendering work.

Shim Layer registers a software interrupt at initialization time for each process, and issue host API invocation message through software interrupt implicitly. This way we don't need to change Android build at all, but implement a Fast Model plugin called SideChannel to capture this software interrupt in Fast Model. All of the parameters and the data pointed to by pointer parameters are packed together as a message and passed to Reconciler in host through SideChannel. The Reconciler completes the followings, (a) Maintain GLES states in host and prepares parameters for calling APIs in host (b) Call API in host through Mali Emulator [12] or native graphics driver (c) Convert frame buffer in host to target format and write back.

For some APIs that need to expose the Open GL state in host, e.g. TransformFeedbackBuffer and ShaderStorageBuffer, we have to fetch the buffer data in host and write them back to the buffer memory returned by graphics driver in target. This could make sure all APIs in target have complete semantics, so the code running on CPU in graphics driver can always work correctly. Mali Emulator is an ARM commercial product which is to build a fake Mali device driver by converting OpenGL ES to Open GL, so it receives the API from Reconciler and dispatches them to real native graphics driver. If native graphics driver in host directly support GLES, this layer would be unnecessary, and we can directly pass the GLES API to host driver. Some APIs like glDraw and glClear will trigger real rendering work in host driver. When API eglSwapBuffer is issued in host side, Reconciler will read pixels from framebuffer in host and write to frame buffer address in target through SideChannel plugin.

In this paper, the solution of using hardware GPU resource to do 3D rendering acceleration is called GGA (Generic Graphics Acceleration), and running API through graphics driver in target on Fast Model is called GRM (Graphics Register Model) execution. We can choose whether we want to turn on GRM or not, and the speed could be much faster without GRM execution. Usually graphics software validation on SoC can start from using GGA without GRM first. If GGA mode can work very well, GRM execution on ARM Fast Model can be turned on to further verify the correctness of driver integration in target.

## 2.2 API Lazy Dispatch

The isolated target and host address space bring troubles in implementing some OpenGL ES APIs on host to simulate the same behavior on target. The challenge is essentially introduced by the incomplete properties of OpenGL ES API parameters. For example, at the moment of passing memory pointer to API, sometimes the size of the memory being passed to OpenGL state machine is not specified, and it won't be determined until it is going to be used by other APIs, so we have to defer the memory copy to the moment this pointer really starts to be used. To support the deferred memory copy we have to introduce an intermediate state on top of GLES state machine in host side, and GLES APIs are dispatched lazily when necessary.

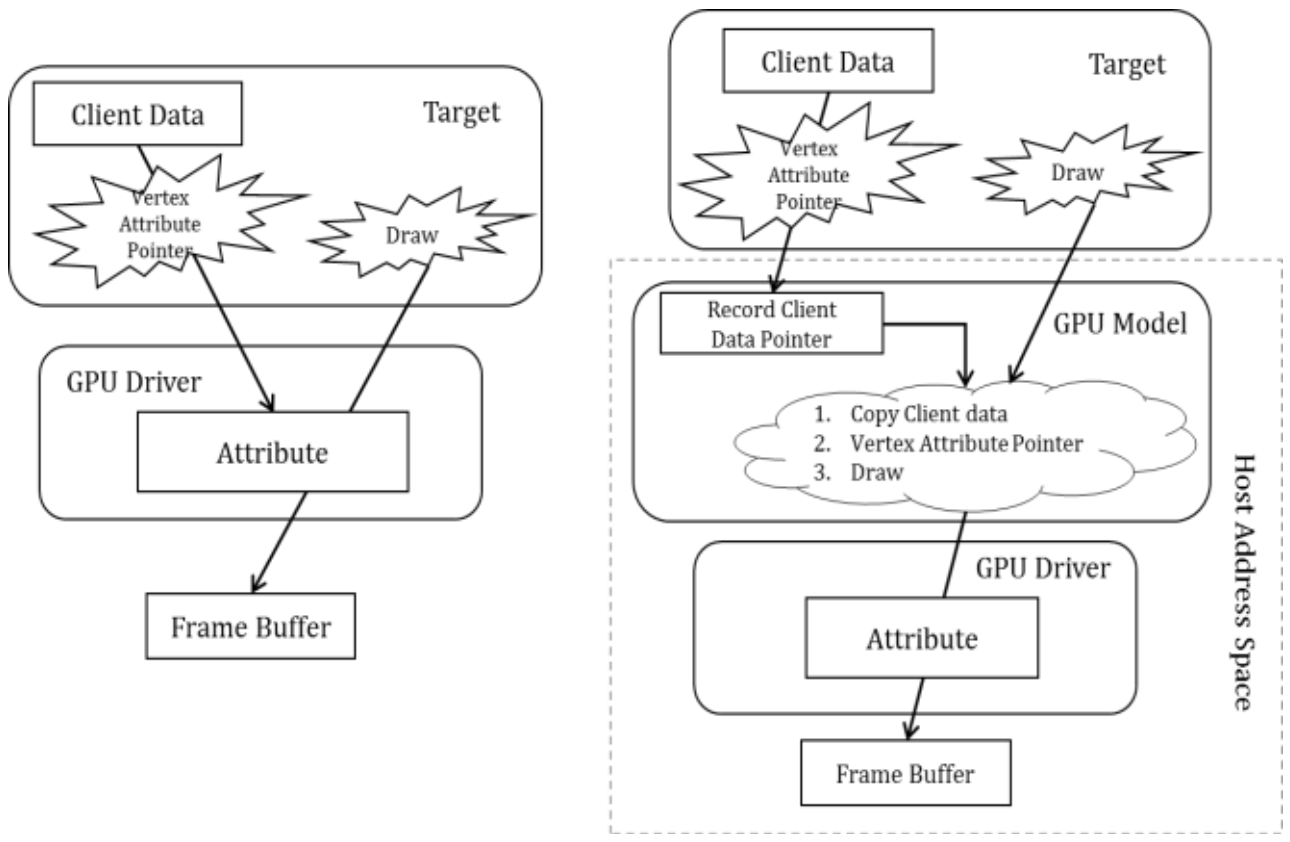

Figure 3. API Lazy Dispatch

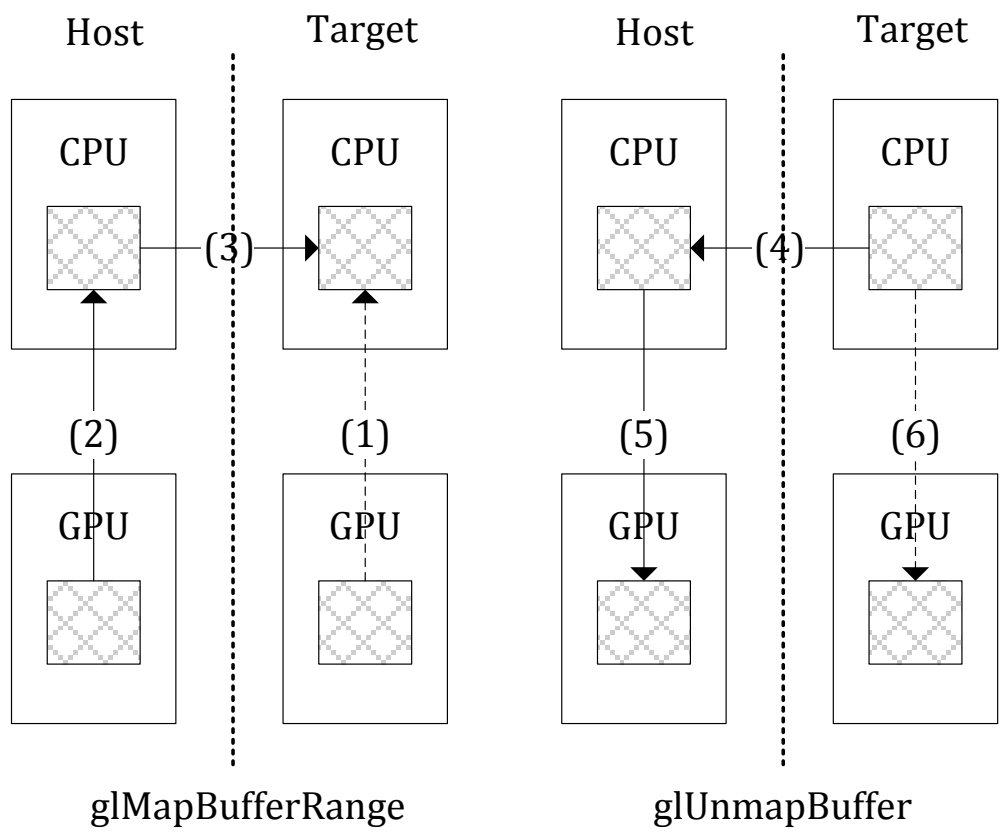

Figure 4. Memory coherence between host and target

In Figure 3, glVertexAtrribPointer specifies the location and data format of the array of generic vertex attributes at index to use for rendering. When it is issued in target side, if we want to get it issued in host side as well, we have to copy all of the CPU accessible memory data specified by parameters from target to host. Unfortunately, OpenGL ES doesn't design such API with specified data size to be used, but it only informs GPU that the memory address in CPU to be accessed, and we won't know exactly the size until glDraw time. It provides a kind of usage flexibility for end user, although it is not a self-contained solution. In this situation, we have to defer the dispatch of glVertexAtrribPointer in host side until glDraw time by using a batch dispatch solution in host. This implies we have to record and maintain some OpenGL ES states in host side to map the corresponding states in target, and then use these states when necessary.

### 2.3 Memory Coherence Between Target and Host

The memory access between CPU and GPU can be guaranteed by operating system and driver support in kernel space naturally. However, this would not be the case when we want to dispatch target API to host in a different address space. This is particularly interesting for two special APIs glMapBufferRange and glUnmapBuffer, which are to map and unmap all or part of the data store of a specified buffer object from GPU to CPU. Figure 4 explains how to handle these two APIs correctly. For glMapBufferRange, we should call in target first, and then call it in host followed by copying mapped buffer data from host to target, i.e. (1)→(2)→(3). For glUnmapBuffer, we should do it in a reversed order, that is, copy the buffer back from target to host followed by calling it in host to unmap the buffer object for host GPU, and at last we call it in target as well, i.e. (4)→(5)→(6). Steps (3) and (6) make sure the correct semantic execution in target application and they are mandatory for maintaining OpenGL ES buffer object state in target driver.

This way, the memory coherence between target and host can be guaranteed. Although these two APIs may introduce GPU bubble due to synchronization between CPU and GPU, there are still some scenarios using them heavily to pinpointing some feedback buffers in GPU. Monitoring this kind of memory access workload caused by these two APIs could be very interesting for GPU performance tuning.

### 2.4 Target Libraries Injection

With proposed GPU model technique in this paper, two libraries need to be installed in Android target to make the communication between target and host work. One is the Shim Layer library as we discussed before in Figure 2, which is to intercept all of EGL/GLES APIs called from applications or Android itself, and it dispatches all of the target API invocations to host. This solution completely follows the standard Android graphics library loading mechanism. The other one is a library to support native windows handler (NWH) abstraction layer (libnwhal.so). We need this layer to expose the internals of the native windows handler object, because end user may directly pass Android native windows buffer pointer to some APIs like glEGLImageTargetTexture2DOES and others etc. Figure 5 shows two libraries are injected into Android target to make GGA and GRM work.

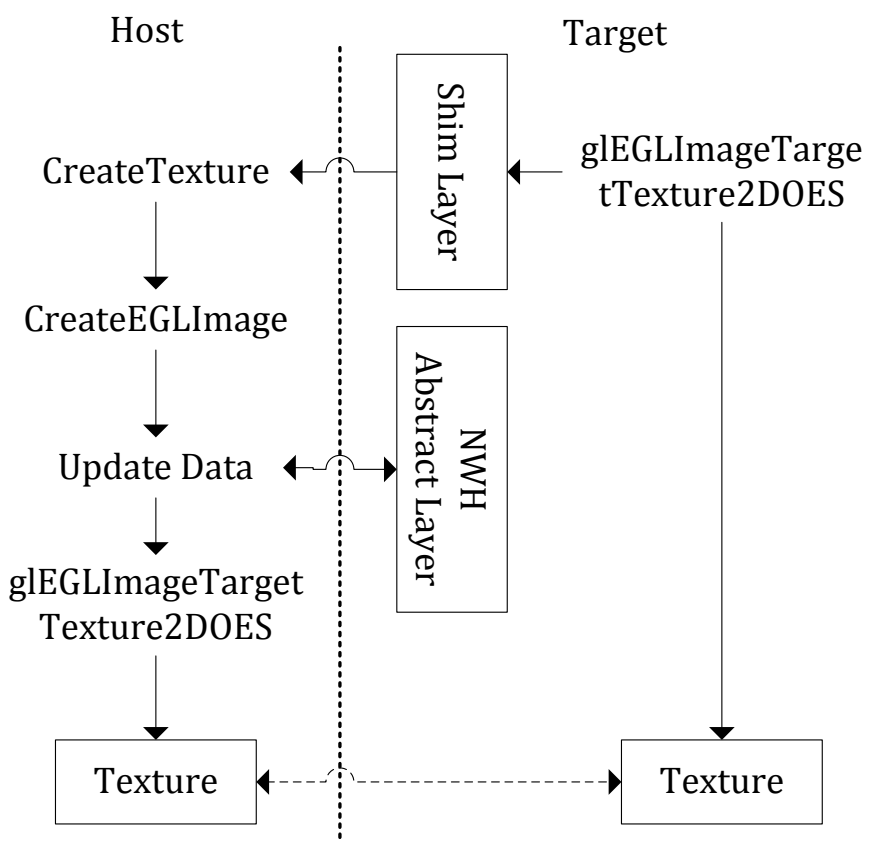

Figure 5. Target library injections

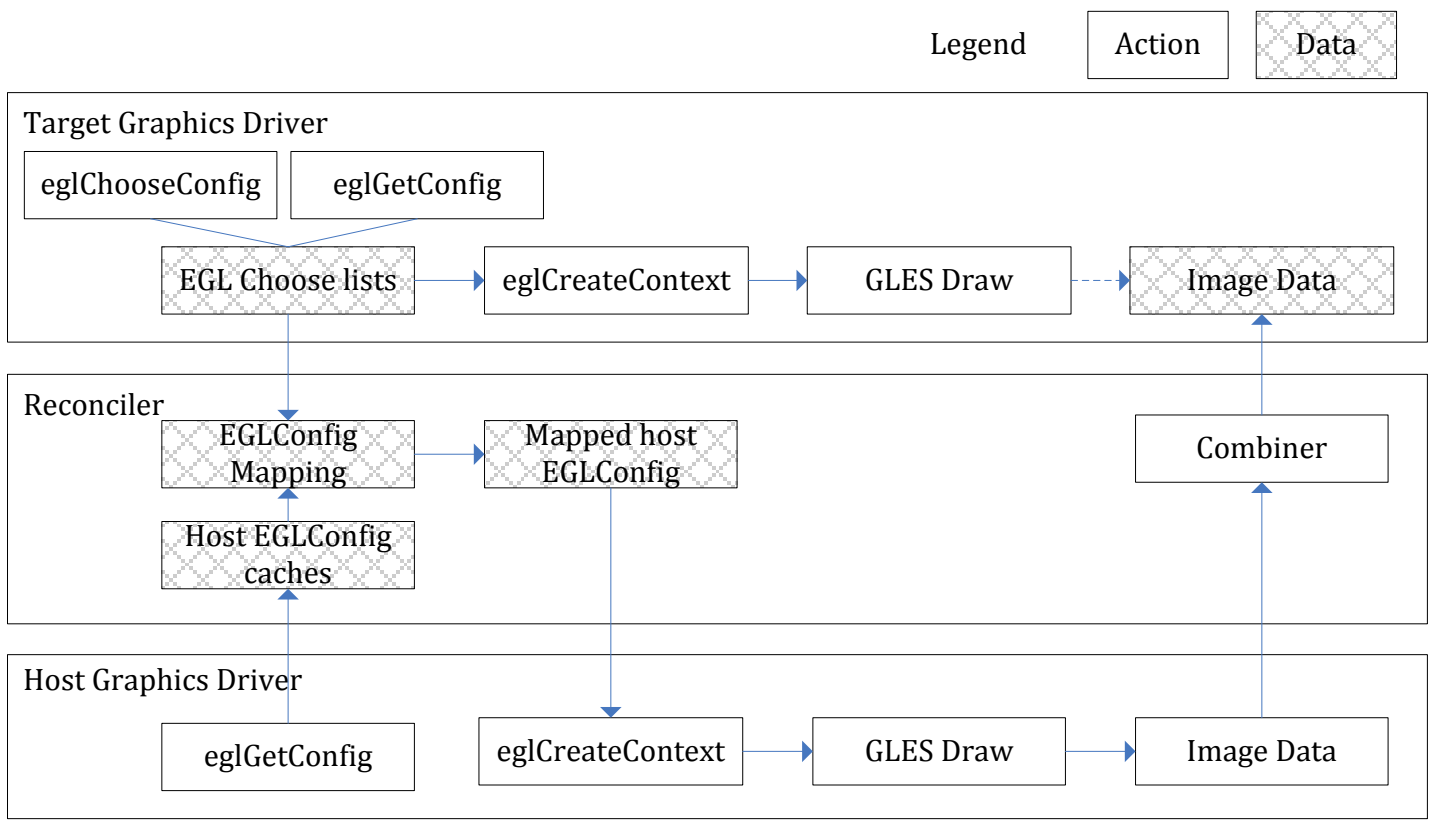

Figure 6. EGLConfig Mapping Flow

An EGL API called eglCreateWindowSurface links two worlds together, i.e. EGL and native windows. EGL is independent of GLES, and it is the interface between OpenGL ES and the underlying native platform window system. Normally end user doesn't need to know the internal of native windows handler in EGL world, and the EGL driver could access the native windows buffer correctly, because it is part of native graphics driver and it can know of Android system. However, in our proposal of accelerating 3D rendering using host API, host side must know the internals of native windows buffer in target to support three special APIs eglCreateWindowSurface, eglCreateImageKHR, and glEGLImageTargetTexture2DOES. For example, API glEGLImageTargetTexture2DOES supports to define an entire 2D texture array from EGLImage for buffer type EGL_NATIVE_BUFFER_ANDROID, which implies host side needs to understand the layout of current native windows handler in target. We have to copy the native windows buffer data from target to host, and build a texture to create a fake EGLImage accommodating the same content in host and eventually to make glEGLImageTargetTexture2DOES work in host side to obtain the texture mapping to the same one in target.

## III. GRAPHICS DRIVER VERIFICATION IN TARGET

With the capability of target libraries injection provided by GGA, verification functions can be freely inserted into various application level calls, whilst GRM makes it possible to inject verification logic into the kernel, driver, and even hardware layers. What's important is the presentation of rendered contents would not be impacted much visually, and the performance hurt caused by adding more and more verification logic can be well controlled. To fully verify the graphics driver as well as system integration, EGL interfaces need to be specially handled due to inconsistency between host and target.

### 3.1 Error Code Checking

From the viewpoint of OpenGL ES API, eglGetError and glGetError are the generic interfaces of checking application level problems at run-time then decide the subsequent code path. However, these would bring somewhat interference to the 3D state machine, hurt performance a bit, and most applications would not frequently do such checking unless very necessary. In GGA and GRM execution, we call egl/glGetError API respectively as soon as a GLES or EGL API is finished in both host and target. With this method we can obtain the error code for all APIs and get them cached in both side, while the implementation of egl/glGetError API itself becomes a little bit tricky by using the value in cache instead. The GRM configurable switch can turn on or off this feature. When it is on, the APIs defined a white list supported by toolset is going to be checked for error code comparison at run-time. However, it could slow down run-time speed a lot, so we can turn it off after finishing the correctness verification. This way application and game engine developers can feel free to do complete or partial error checking anytime, without changing the application source code. This can also be a suitable tool for GPU driver debugging on the condition that 3D applications are extremely complicated. This error code checking feature is covered by the verification module as shown in Figure 1.

### 3.2 Render Buffer Target Analysis

Different Android graphics components are glued by a BufferQueue, in which some buffers are allocated by Android gralloc module. In Android target, each window surface has the following procedure inside native graphics driver sequentially, (a) Dequeue a piece of buffers from BufferQueue at the very beginning of glDraw inside target driver. (b) glDraw to make all 3D rendering actions work, and write rendering result to the buffer we just dequeued. (c) Enqueue the renderer buffer into BufferQueue by eglSwapBuffer. Writing the rendering result back to the buffer in target requires us to know the address of the buffer dequeued by glDraw at step (a). Unfortunately, this address is usually hidden in graphics driver, and won't be exposed to end user. Without knowing this, we would have to use DequeueBuffer and CancelBuffer to obtain the buffer address, but the prerequisite is the buffer cancelled could be dequeued out of BufferQueue shortly by the following glDraw.

However, in Fast Model we still have chances to know this address by analyzing RenderTargetDescriptor, which is issued by graphics driver to graphics hardware. This special analysis action is implemented inside GRM, and all of the instructions issued to GPU by graphics driver are visible inside GRM, so we intercept the register write into the graphics device to figure out the rendering target buffer address by accessing the memory address in CPU through MMU as a master device.

In some other situations, if we don't have chances to parse RenderTargetDescriptor, an alternative solution is to use an EGL extension called EGL_KHR_lock_surface3, with which the color buffer of EGL surface can be mapped to user space, then we can write rendering result into this exposed location. However, this may not be a good choice because the cost of implementing this extension in graphics driver is breaking the GPU rendering pipeline, so the performance could be hurt accordingly.

### 3.3 Target Extension Support

In order to have the well-balanced 3D resources management in consideration of quality vs. performance on mobile systems, GPU driver generally provides a number of special 3D API extensions on mobile systems. For example, EGL_ANDROID_image_crop is to ask driver to process 3D image or texture resources cropping rather than using the texture transform matrix, which can alleviate computation in some cases. For pure software emulation, implementing these special interfaces can be an overhead and applications may adaptively change the inner logic if cannot find specific extension support, while GGA can help capture the extensions and forward to host side implementations, which can provide a set of unified inputs and results for keeping target side applications go through the same code path in comparison with running on real devices.

### 3.4 EGLConfig Mapping

If GRM execution is turned on, target and host graphic driver can present different features. This means some features which can't be supported on both side should be dropped or converted into alternative implementations. However, in order to verify the software stack in Android, we need to keep the features of target graphic driver as much as possible. The EGLConfig mapping is one of the most critical issues for rendering. Target graphic driver can support various kinds of EGLConfig, e.g. RGB565 plus a lot of other features, but some host graphic drivers don't support them at all. When a target EGLConfig is mapped to a host one, we have to make sure the rendering results in host match the same capability on target side. Otherwise, we need to do simulation. For example, if application in target is asking for a format RGB565, which host side can't support, the combiner inside Reconciler module can resize the framebuffer data for target side, such as converting the host side 4-byte RGBA8888 image data into target side 2-byte RGB565 image data.

Figure 6 describes the EGLConfig mapping flow. First, the host side EGLConfigs are all cached when eglInitialize is called. Second, the target graphic driver generates the EGLConfig lists for eglChooseConfig and eglGetConfig, while each EGLConfig is mapped to a host side EGLConfig that best supports target. Third, the mapped host EGLConfig is used for host side GLES draw, and the final image data is reorganized by combiner and then written back to target side framebuffer.

## IV. REMOTE RENDERING ACCELERATION

The SoC simulation using software is very slow traditionally, although the generated platforms using ARM Fast Model can reach simulation speed of ranges between 100-500 MIPS on a typical workstation. A lot of SoC development vendors are using cluster with thousands of powerful machines to speed up the verification work. The solution we proposed in chapter 3 requires to install graphic cards on all of the machines running Fast Model. It becomes a great challenge to enable such a lot of graphic cards on huge number of machines in cluster in reality. Therefore, we explored a new solution by remotely accelerating graphic computation related to 3D rendering in cluster over network.

The initial experiment showed the rendering speed of ARM Mali SDK example Cube for GGA is ~4FPS, which can vary depending on the different hardware configurations, so it implies the memory buffer communication speed between host and target on a display with 1024x768 resolutions display is at least ~100Mb/s. Furthermore, optimization using lossless compression and rendering buffer cache in host for each window surface in target could reduce the memory data transition a lot. This is not a problem at all for giga band Ethernet, so we can eventually use a single remote GPU resource to accelerate multiple 3D rendering verification instances in cluster.

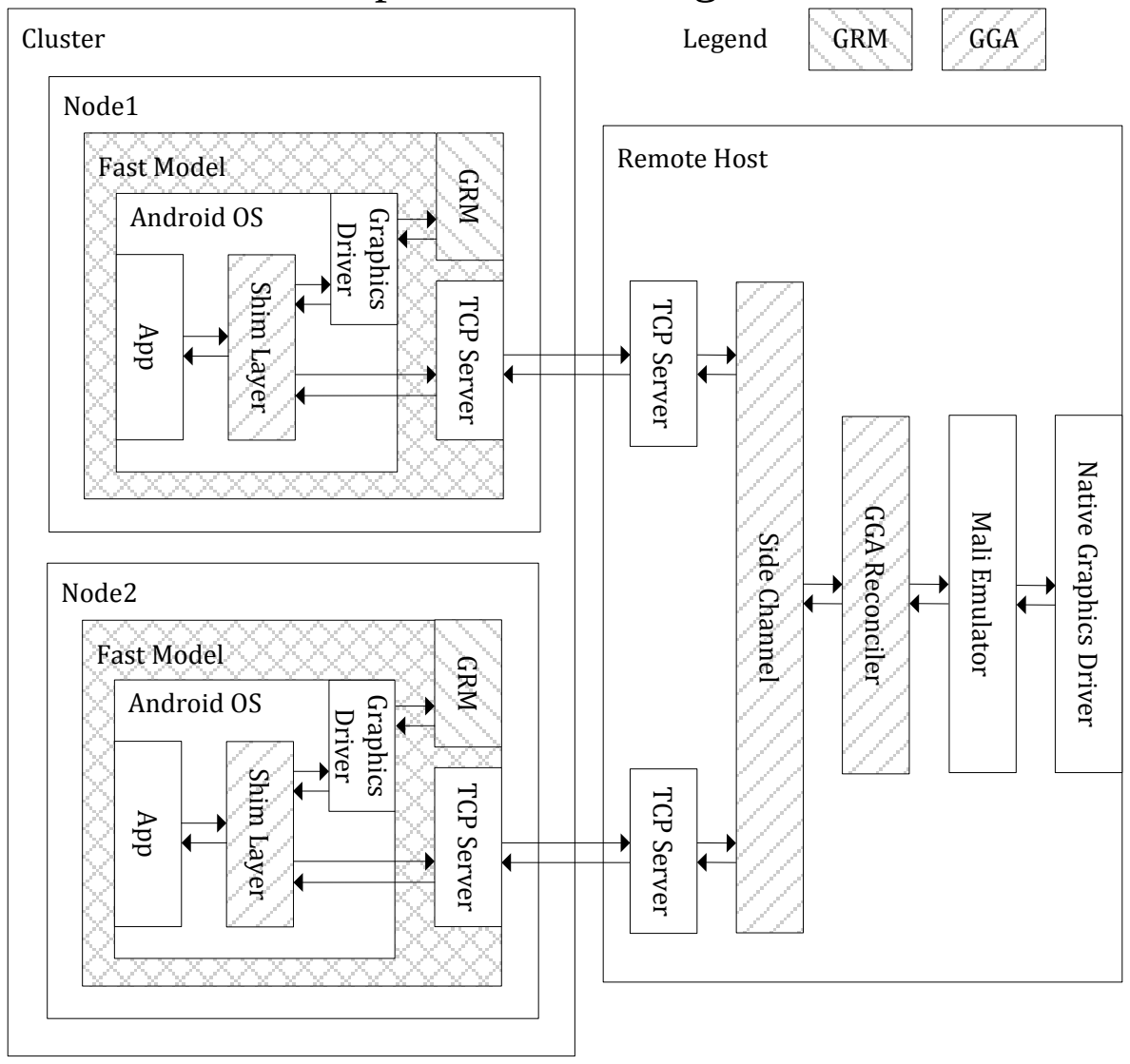

Figure 7. Remote Rendering Acceleration

Figure 7 explains how remote rendering acceleration work in ARM Fast Model environment. First, APIs issued by application running in Android target directly go into Shim Layer. Second, these APIs can be dispatched to target graphics driver and go into GRM, or they are passed to remote GGA reconciler via Fast Model SideChannel plugin, in which remote IPC is supported. Third, in host APIs are eventually passed to host graphics driver via Mali Emulator. And the rendering result will be transferred back to target in reversed direction at appropriate moment, e.g at eglSwapBuffer invocation time, or it will be fetched by GRM at a moment triggered by VSYNC signal from display device. We will discuss more about eglSwapBuffer in chapter 4.2.

### 4.1 Batch Mode

To further reduce the networking traffic between accelerator and the cluster, a batch mode for OpenGL ES API is introduced. The GLES API can be accumulated and issued all together until we come across two situations, (a) Return memory by pointer passed by parameter. (b) eglGetError or glGetError is called and needs to be returned to application. In reality, there are tons of situation (b) cases for running real benchmarks. Most of the time the normal applications like commercial benchmarks don’t return abnormal error code, so we can bypass it by always returning no error. With this solution, a lot of the API passing traffic is reduced between host and target.

### 4.2 Frame Buffer Identification Analysis

As we discussed before, the host side GLES API execution must depend on the memory buffer copied from target. Actually the biggest memory bandwidth consumption between host and target is the native rendering buffer read from host and written to target for each window that is visible. Figure 8.a shows how rendering buffer is read and written with regard to Surface Flinger doing hardware composition. Client is to describe the target running on Fast Model, while server is the remote host running real 3D rendering work. When eglCreateSurface is called in client, a native buffer is created. In the meantime, we inform server to create a corresponding buffer for GPU in server. When rendering in server side is finished, eglSwapBuffer will be invoked to copy memory back to client as indicated by (1) and (2). In the chart we use two buffers B0 and B1 as examples. When Surface Flinger is triggered by VSYNC signal from display device, the software composition operation will dispatch GLES API glEGLImageTargetTexture2DOES to copy the buffers B0 and B1 in client to B0’ and B1’ in server. At this moment B0’ is B0, and B1’ is B1 in terms of data inside. After the real composition operation is completed in server, a frame buffer fitting display size will be generated, and eglSwapBuffer is invoked to copy it back to framebuffer device fb0 in Android OS of client. This procedure is the standard behavior if we dispatch all GLES API calls from client to server without optimizations.

In fact, the memory copies as described by (1)(2)(3)(4) can be saved, and the huge memory bandwidth consumption can be reduced accordingly by reducing unnecessary memory copies. The optimization is described as Figure 8.b. In order to recognize B0’ is identically B0, and B1’ is identically B1 in server side, we check the display size in target as a client. If the size of the rendering buffer is just the size being used by the display we assume the rendering buffer is for hardware composition target, i.e. device fb0. With this solution, about ¾ of frame buffer copy could be reduced. The Figure 9.a shows network bandwidth before optimization, so we have almost the same number for both upstream and downstream. After optimization we have Figure 9.b, and it shows 75% of upstream data is reduced. This solution can cover most of the real cases, although it has defects for some corner cases. We will discuss it further in chapter 6.2.

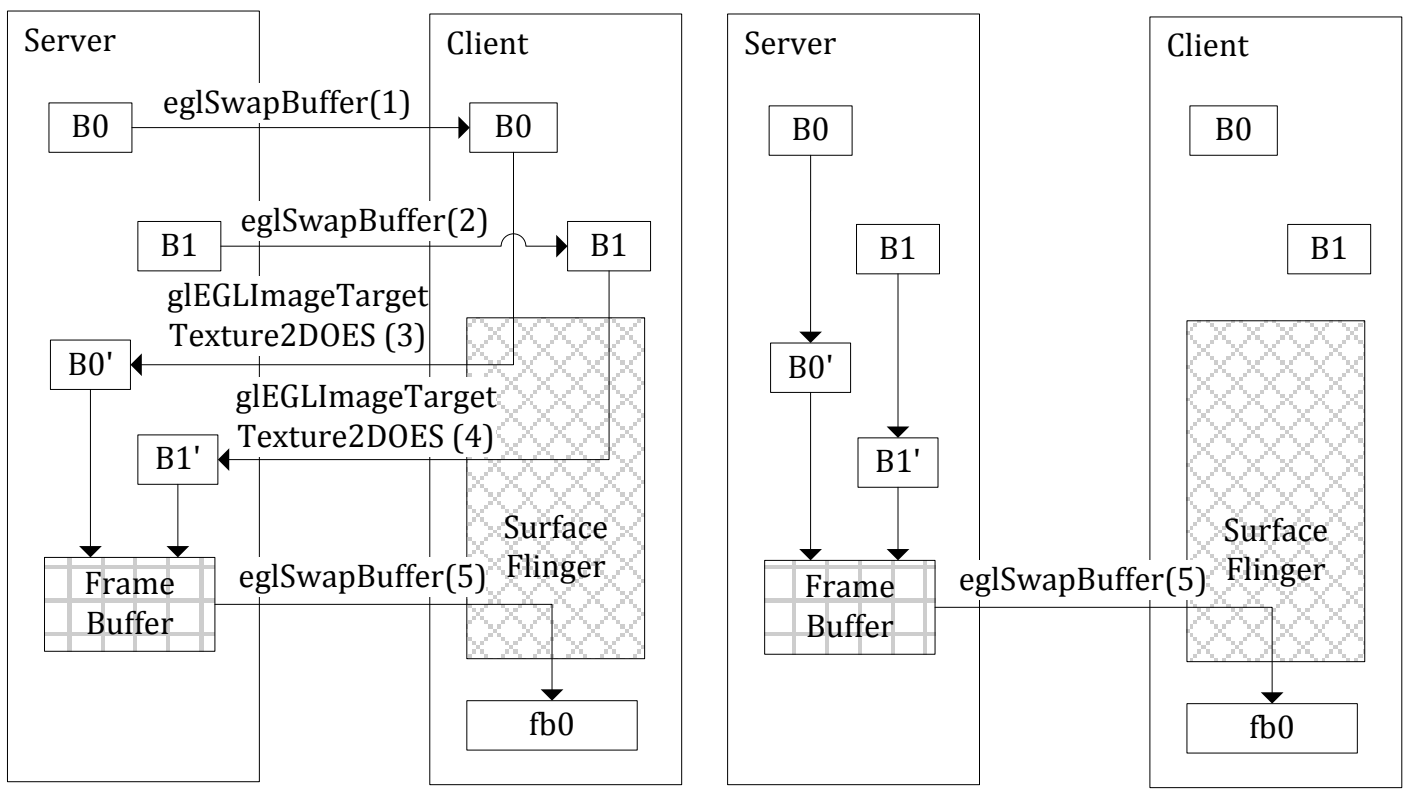

a. Before Optimization b. After Optimization

Figure 8. Rendering Buffer Access

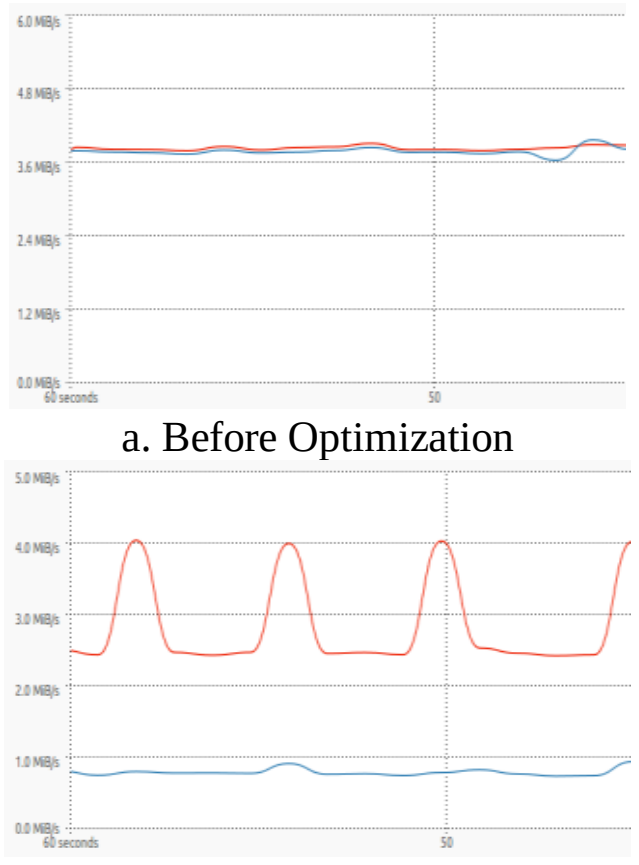

a. Before Optimization

b. After optimization

Figure 9. Remote Modelling Network Traffic

## V. Experiment and Performance

This section presents a set of experiment results aimed at demonstrating the 3D rendering acceleration effectiveness for graphics application running on Fast Model. All experiments run on Intel(R) Xeon(R) CPU E5-2690 v4 @ 2.60GHz. Experimental evaluations were performed using commercial GPU performance benchmarks

Antutu, GFX, and Unity. Android boot up is also measured to compare the effect of software rendering provided by google. For all of these benchmarks we can view the run-time execution result in Android GUI.

### 5.1 Performance comparison

Figure 10 gives performance comparison among software rendering, GGA and GRM executions. Software rendering can't support real OpenGL ES APIs to run most commercial benchmarks, and it can only make android boot-up successfully.

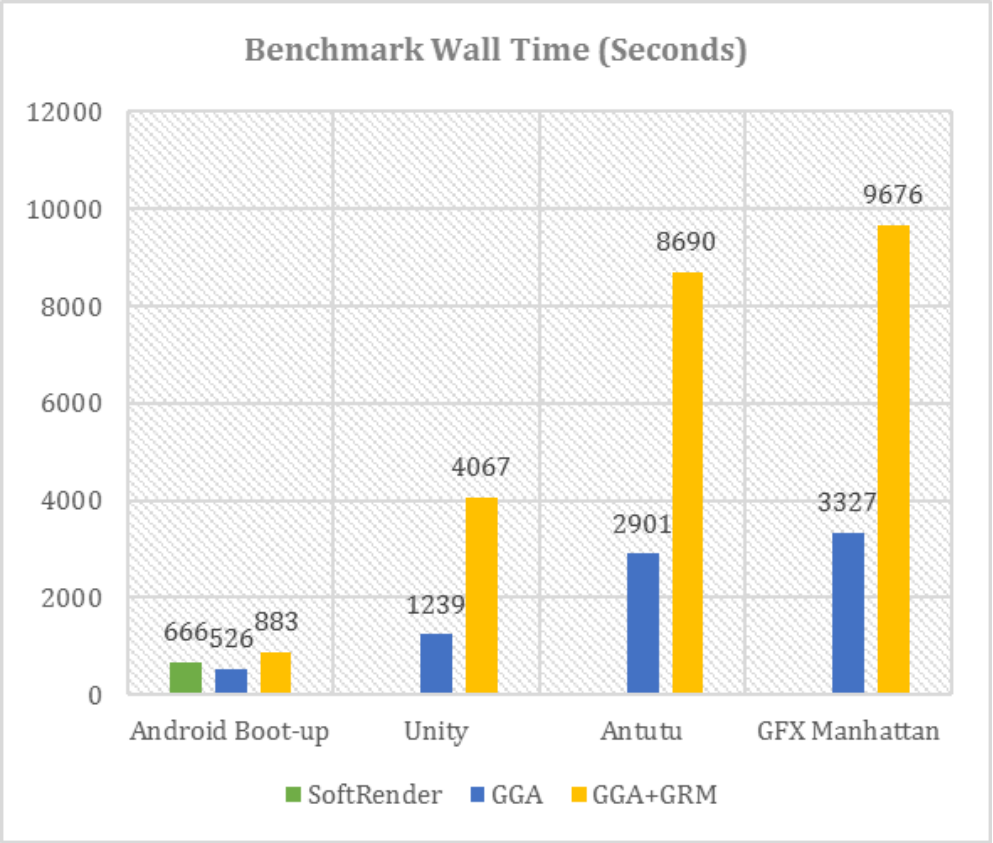

Figure 10. Benchmark Wall Time Comparison

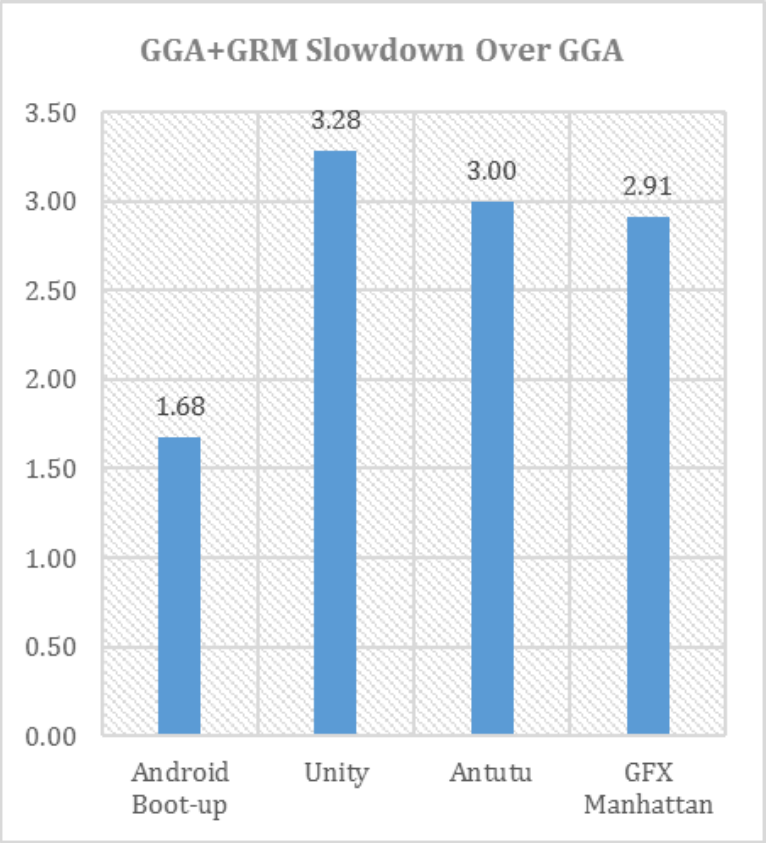

Figure 11. GGA+GRM Slowdown

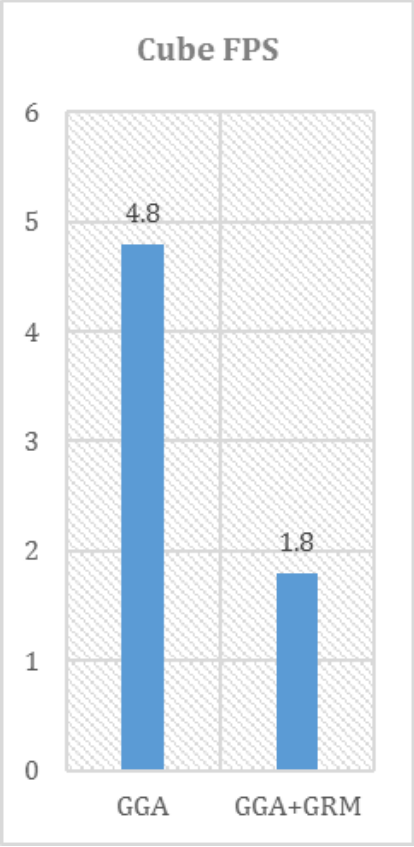

Figure 12. Cube FPS

Figure 11 shows how much GRM execution slow down Fast Model for running real benchmarks. The mean value for these 4 typical benchmarks is 2.72. GRM execution doesn't change GGA behavior for 3D rendering, and there isn't new overhead added for GRM, so the slowdown is largely from the graphics driver execution in target as well as the relevant behavior changes in Android surface flinger. Figure 12 gives FPS (Frame Per Second) number for running Mali Cube example, which is from Mali graphics SDK. The FPS slowdown ratio is 2.67 that is a number very close to the wall time slowdown ratio.

### 5.2 Performance bottle neck for GRM

Figure 13 gives break-down of the EGL/GLES execution time for Mali Cube. Figure 13.a lists all APIs taking >1% execution time, in which EGL API eglSwapBuffers has the longest execution time, and we can see from 13.c that the driver in target takes the longest time to execute this API. This is largely because eglSwapBuffers performs an implicit flush operation on the context and it introduces huge memory copy pressure to Fast Model. Figure 13.b shows most of the execution time is in graphics driver API execution in target.

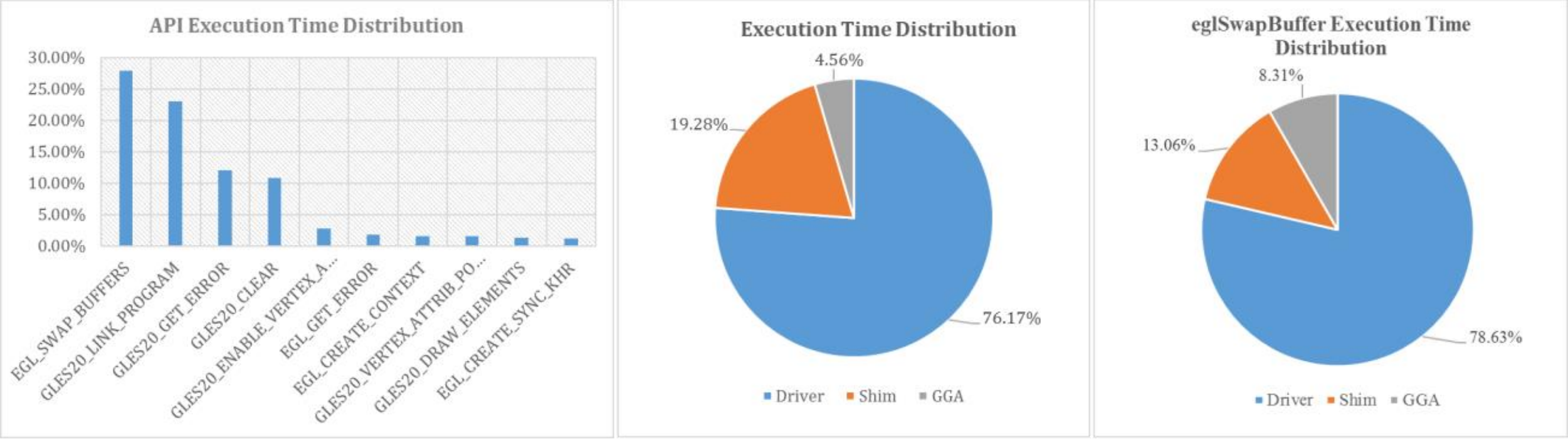

a. APIs break-down b. Break-down based on module c. eglSwapBuffer break-down

Figure 13. API Execution Time Distribution

## VI. CONCLUSIONS AND FUTURE WORK

The work described in this paper is based on graphics model development of ARM Fast Model product. The 3D rendering acceleration using resources in host significantly improves graphics software stack validation speed, no matter whether the resource is local or remote. Using the technology proposed in this paper, we can run real benchmarks like Antutu, GFX, and Unity to prove the software stack integrity without changing anything in native graphics driver and Android OS build. Therefore, the image that has been verified on Fast Model can be directly further verified on real chip without any changes. Comparing to the software rendering provided by Android open source code base, the graphics model of ARM Fast Model breaks the ice of validating graphics technology for SoC due to high usability, high performance and high efficiency. Besides, the proposed technology can also help the graphics research by exposing the memory traffic between GPU and CPU. Apart from these advantages, we also look forward to solving some potential problems in the future.

**6.1 Direct Memory Write**

The most challenging problem in our current solution is huge memory copy between target and host. We have to do this because target and host are in completely different address space. A normal path of transferring memory is to use Fast Model Side Channel plugin, which essentially is a back door of target world communicating with host world. Before writing to target, we touch the memory being written to make sure the page is really in memory in target, and after doing that the memory write passed by host needs to go through the direct memory access bus simulated by Fast Model, and eventually the render buffer memory is written into physical frame buffer in target world. This is a very long execution trace in practice, which slows down Fast Model execution. Since Fast Model subsystem can know all the devices being supported, it is possible to expose the physical frame buffer address through a special back door in host side. But the problem is the size of the memory pointed by this address cannot be larger than a single page, that is, we can only guarantee the continuity within a page of the target memory being shared with host. If we can always guarantee the continuity crossing pages, we would be able to directly write rendering result into the shared memory in target to avoid the detour of memory write and performance would improve a lot.

**6.2 Frame Buffer Identity Detection**

The solution described by chapter 4.2 may not always work. For some corner cases, if the rendering buffer size is the same as the display size in target, but the buffer is not the target of surface hardware composition, the optimization would not work any longer. How can we always correctly know B0’ is B0, and B1’ is B1 in server side? In figure 8.a, we can’t see explicit connection between B0/B1 and B0’/B1’ in server side, but they are implicitly connected by the native buffer B0/B1 in client side. Therefore, we need a solution to inform server the connection or we need to infer this connection using the characteristics of the frame buffer. Unfortunately, there isn’t a perfect solution to know B0/B1 are the same one as B0’/B1’ in user space. To solve this issue, we have to build and install a small fd query driver in Android OS to check whether B0/B1 and B0’/B1’ can be mapped with each other or not. In this driver, we can find the hidden data structure in driver through the fd descriptor provided in user space, and then it would be easy for us to know the relationships of the buffers exposed in user space.

In figure 14, the dot lines are to query the unique kernel buffer ID through the driver installed in target, so we will be able to use the unique ID to check the relations of the buffer allocated in target, and then decide if we really need to copy the native buffer from client to server. If we find there is already a buffer cache in server, we will not do that, and accordingly we don’t need to copy the buffer back from server to client either. All of the rendering work in server will be based on the buffer cache, and we only write back the frame buffer specially mapped to device fb0 in client side. Therefore, a lot of memory bandwidth will be saved.

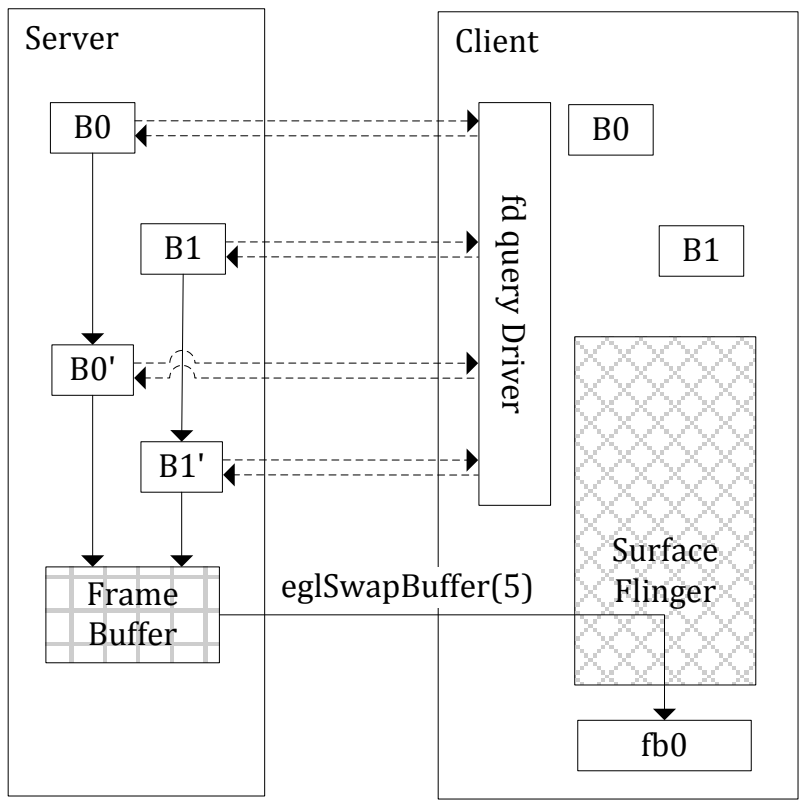

Figure 14. Frame Buffer Identity Detection

**6.3 Multimedia Development and Test Support**

There are several popular video players for Android as well as several content streaming services such as Netflix available for Android. Most of these video playback applications use Android Java MediaPlayer APIs supported by Stagefright which is able to off-load video decoding to dedicated hardware blocks through OpenMAX IL (Integration Layer), and performs session management, time-synchronized rendering, and DRM etc. GGA can speed up the rendering work much, in comparison with the built-in GLES CM 1.0 software rendering on QEMU. More importantly, vendors usually design their own hardware codec by using the OpenMAX IL standard, which provides a standardized way for Stagefright to recognize and use custom hardware-based multimedia codecs. The proposed methodology can help verify the integrations with Stagefright as a whole in the CSS (Compute Subsystem), and better support the automated testing harness provided by Android Media CTS (Compatibility Test Suite).

### 6.4 Turn On GRM On-the-fly

GRM execution always slows down the rendering acceleration because we have to synchronize with the API execution on model. The pure GGA mode is the up-limit speed for graphics software stack verification in Fast Model. It would be much better if we can dynamically turn on or off GRM execution, and end users only need to turn on GRM for the execution in which they have interest. If we turn it on at a random moment, graphics driver could miss out on certain API calls then the state within the driver and GGA in host will start to disagree, e.g. trivially if the driver doesn't see a glGenTexture call, but later sees a glTexImage2D call then it will fault it. However, it is still possible to only turn on or off GRM execution at the boundary of a process live range. Usually different processes in target have initializations to create different GLES contexts, so we can easily make differences in both target and host to dynamically decide if GRM needs to be executed in terms of process ID. The GLES state wouldn't be messed up if only we guarantee the rendering completeness within a single context.

### 6.5 Gralloc-ION Related Future Work

There are many hardware blocks on a typical SoC running Android, where buffers need to be managed and shared between drivers and processes. Gralloc and Android ION framework define the guideline so the buffer sharing among various kernel drivers can be customized freely on the basis of this. GGA and GRM execution for GPU can make these multimedia kernel drivers happy and test the software pipeline more systematically, then support the full system verifications in Fast Model. Furthermore, Android fence based synchronization mechanism is the key of coordinating the cooperation of media devices, including video decoder, display controller, any hardware being used by HWComposer, and the GPU driver as well. GGA and GRM execution can help pinpoint key places of doing such synchronizations from viewpoint of both user and kernel space. This would bring great value for vendors who what to integrate multiple media devices, or tune cross-device technologies, e.g. utilize ARM Frame Buffer Compression (AFBC) protocol to minimize the amount of data transferred between IP blocks within a SoC.

## VII. Acknowledgement

The work described in this paper was supported by Fast Model project sponsored by ARM Ltd. The authors would also like to thank Jesse Zhou (ARM) for his advices and feedback on this work.